\documentclass[10pt,a4paper]{article}
\ifx\pdfpageheight\undefined\PassOptionsToPackage{dvips}{graphicx}\else%
\PassOptionsToPackage{pdftex}{graphicx}
\PassOptionsToPackage{pdftex}{color}
\fi
\newcommand{\type}{\mathsf{type}}

\newcommand{\HOL}{\lambda\mathsf{HOL}}
\newcommand{\U}{\lambda\mathsf{U}}
\newcommand{\Pos}{\square}%{\scalebox{1.4}{$\square$}}

\usepackage{color}

\newcommand{\EMP}[1]{\emph{\textcolor{red}{#1}}}

\usepackage[all]{xy}
\usepackage{url}
\usepackage{verbatim}
\usepackage{latexsym}
\usepackage{amssymb,amstext,amsmath}
\usepackage{tikz-cd}
\usepackage{epsf}
\usepackage{epsfig}
\usepackage{a4wide}
\usepackage{verbatim}
\usepackage{proof}
\usepackage{latexsym}
\usepackage{amssymb}
\newtheorem{theorem}{Theorem}[section]

\newcommand{\intro}{\mathsf{intro}}
\newcommand{\match}{\mathsf{match}}

\usepackage{float}
\floatstyle{boxed}
\restylefloat{figure}

\def\oge{\leavevmode\raise
.3ex\hbox{$\scriptscriptstyle\langle\!\langle\,$}}
\def\feg{\leavevmode\raise
.3ex\hbox{$\scriptscriptstyle\,\rangle\!\rangle$}}

\def\N0{\hbox{\sf N}_0}

\newcommand{\Prop}{\mathsf{*}}

\newcommand{\Pow}{\mathsf{Pow}}

\begin{document}

\title{A variation of Reynolds-Hurkens Paradox}

\author{Thierry Coquand, University of Gothenburg, Sweden}
\date{}
\maketitle

%\rightfooter{}

\section*{Introduction}

We present a variation of Hurkens paradox \cite{Hurkens95}, itself being a variation
of Reynolds ``paradox'' \cite{Reynolds84}, as used in \cite{coq94-2}. We first explain
a related paradox in higher order logic, which can be seen as a variation of Russell's paradox.
We then show how this paradox can be formulated in system $\U^{-}$. We finally argue that an analysis of the
computational behavior of this paradox requires to extend existing type systems with a first class
notion of definitions and head linear reductions, as advocated by N.G. de Bruijn \cite{deBruijn87}.

\section{Some paradoxes in minimal Higher-Order logic}

We first present some paradoxes in some extensions of the system $\HOL$, minimal Higher-Order logic, 
described in \cite{geuvers07}. This system can be seen as a minimal logic version of higher-order logic
introduced by A. Church \cite{Church40}. With the notation of \cite{geuvers07}, it has sorts $\Prop,\Pos,\Delta$
with $\Prop:\Pos$ and $\Pos:\Delta$ and the rules
$$
(\Prop,\Prop),~(\Pos,\Pos),~(\Pos,\Prop)
$$
We denote by $X,Y,\dots$ types of this system.

We can define $\Pow:\Pos\rightarrow\Pos$ by $\Pow~X = X\rightarrow\Prop$
and $T:\Pos\rightarrow\Pos$ by $T~X = \Pow~(\Pow~X)$.

 Note that $T$ defines a {\em judgmental} functor: if $f:X\rightarrow Y$ we can define
 $T~f:T~X\rightarrow T~Y$ by
 $$
T~f~F~q = F~(\lambda_{x:X}q~(f~x))
 $$
and we also have if furthermore $g:Y\rightarrow Z$ the judgemental equality (here $\beta$-conversion \cite{geuvers07})
$T~(g\circ f) = (T~g)\circ (T~f)$ defining $g\circ f$ as $\lambda_{x:X}g~(f~x)$.

 We assume in this section to have a type $A:\Pos$ together with two maps
$\intro:T~A\rightarrow A$ and $\match:A\rightarrow T~A$.

We explain now how to derive simple paradoxes assuming some convertibility properties of these maps.

\subsection{A variation of Russell's paradox}

The first version is obtained by assuming that we have $\match~(\intro~u)$ convertible to $u$, i.e. $T~A$
is a judgemental retract of $A$.

Intuitively, we expect $\Pow~A$ to be a retract of $T~A$, and this would imply that $\Pow~A$ is a retract of $A$
and we should be able to deduce a contradition by Russell's paradox. One issue with this argument is that it holds
only using some form of \EMP{extensional} equalities, and we work in an intensional setting. One way to solve this
issue is to work with Partial Equivalence Relations; this is what was done in \cite{coq94-2}. The work
\cite{Hurkens95}, suggests that there should be a more direct way to express this idea, and this is what we present here.

The contradiction is obtained as follows. We first define a relation $C:\Pow~A\rightarrow\Pow~A$
$$
C~p~x~=~p~x\rightarrow \neg(\match~x~p)
$$
where, as usual, we define $\perp:\Prop$ by $\perp = \forall_{p:\Prop}p$ and
$\neg : \Prop\rightarrow \Prop$ by $\neg~p = p\rightarrow \perp$.
We can then define $p_0:\Pow~A$
$$
p_0~x~=~\forall_{p:\Pow~A}C~p~x
$$
We can also define $X_0:T~A$
$$
X_0~p~=~\forall_{x:A}C~p~x
$$
and $x_0:A$ as $x_0 = \intro~X_0$. We can then build $l_1:X_0~p_0 = \match~x_0~p_0$
$$
l_1~x~h~=~h~p_0~h
$$
and $l_2:p_0~x_0$ by
$$
l_2~p~h~h_1~=~h_1~x_0~h~h_1
$$
But this is a contradiction since $\match~x_0 = \match~(\intro~X_0) = X_0$
by hypothesis, and hence $l_2~p_0~l_2~l_1$ is of type $\perp$.

\medskip

We can summarize this discussion as follows.

\begin{theorem}\label{simpl}
  In $\HOL$, we cannot have a type $A$ such that $\Pow~(\Pow~A)$ is a judgemental retract of $A$.
\end{theorem}

This can be seen as a variation of Russell/Cantor's paradox, which states that $\Pow~A$ cannot be a retract of $A$. Here we
state that $T~A$ cannot be a retract of $A$.

 \subsection{A refinement}

We define $\delta:A\rightarrow A$ by $\delta = \intro\circ\match$ and assume
the judgemental equality
$$\match\circ\intro = T~\delta\eqno{(1)}$$
which implies $\match~(\delta~x)~p = \match~x~(p\circ\delta)$.

We now (re)define
$p_0:\Pow~A$
$$
p_0~x~=~\forall_{p:\Pow~A}~p~(\delta~x)\rightarrow\neg (\match~x~p)
$$
and $X_0:T~A$ as before
$$
X_0~p~=~\forall_{x:A}~p~x\rightarrow\neg (\match~x~p)
$$
and $x_0:A$ as $x_0 = \intro~X_0$. Using the judgemental equality $(1)$, it is possible to build
$$s_1:\forall_x~p_0~x\rightarrow p_0~(\delta~x)~~~~~~~~~~~~s_2:\forall_p~X_0~p\rightarrow X_0~(p\circ \delta)$$
by $s_1~x~h~p ~=~ h~(p\circ\delta)$ and $s_2~p~h~x ~=~ h~(\delta~x)$.

We can now define
and $l_0 : \forall_{p:\Pow~A}~p~x_0\rightarrow \neg (X_0~p)$ by
$$l_0 ~p~h~h_0 ~=~h_0~x_0~h~(s_2~p~h_0)$$
using $(1)$ and $l_1:X_0~p_0$ by
$$l_1~x~h  ~=~h~p_0~(s_1~x~h)$$
and $l_2 : p_0~x_0$ by $l_2~p = l_0~(p\circ\delta)$.

For this, we use the judgemental equality $\match~(\delta~x)~p = \match~x~(p\circ\delta)$, consequence of $(1)$.

We can then form the term $l_0~x_0~l_2~l_1$ which is of type $\perp$.

We thus get the following result, using $T~X = \Pow~(\Pow~X)$.

\begin{theorem}\label{refined}
  In $\HOL$, we cannot have a type $A$ with two maps $\intro:T~A\rightarrow A$ and
  $\match:A\rightarrow T~A$ with $\match\circ\intro$ convertible to $T~(\intro\circ\match)$.
\end{theorem}

\section{An encoding in $\U^{-}$}

\subsection{Weak representation of data type}
  
Using the notations of \cite{geuvers07} the system $\U^{-}$ has also sorts $\Prop,\Pos,\Delta$
with $\Prop:\Pos$ and $\Pos:\Delta$ and the rules
$$
(\Prop,\Prop),~(\Pos,\Pos),~(\Pos,\Prop),~(\Delta,\Pos)
$$
We explain in this section why the refined paradox has a direct encoding
in the system $\U^{-}$.

As before, $T$ defines a judgemental functor: if $f:X\rightarrow Y$ we can define
 $T~f:T~X\rightarrow T~Y$ by
 $$
T~f~F~q = F~(\lambda_{x:X}q~(f~x))
 $$
and we also have if furthermore $g:Y\rightarrow Z$ the judgemental equality
$T~(g\circ f) = (T~g)\circ (T~f)$ defining $g\circ f$ as $\lambda_{x:X}g~(f~x)$.

A $T$-algebra is a type $X:\Pos$ together with a map $f:T~X\rightarrow X$.

Following Reynolds \cite{Reynolds84,ReynoldsP93}, we represent $A:\Pos$ by
$$A = \Pi_{X:\Pos}(T~X\rightarrow X)\rightarrow X$$

It can be seen as a weak representation of a data type. If we have $X:\Pos$ and $f:T~X\rightarrow X$
we can define $\iota~f : A\rightarrow X$ by $\iota~f~a = a~X~f$. We can then define
$\intro:T~A\rightarrow A$ by $\intro~u~X~f = f~(T~(\iota~f))$, and we have the conversion
$$(\iota~ f)\circ\intro = f\circ (T~(\iota~f))\eqno{(2)}$$

This expresses that the following diagram commutes strictly

\medskip
\begin{center}
\begin{tikzcd}
  T~ A \arrow[dd, right, "\intro"] \arrow[rr, left, "T~ (\iota~f)"] & & T~ X \arrow[dd, "f"] \\
  \\
  A \arrow[rr, right, "(\iota~f)"] & & X
\end{tikzcd}
\end{center}
\medskip

So $A,~\intro$ represents a \EMP{weak} initial $T$-algebra.

We define next $\match : A\rightarrow T~A$ by $\match = \iota~(T~\intro)$.
Using the conversion $(2)$, we have
$$\match\circ \intro = (T~\intro)\circ (T~\match) = T~(\intro\circ\match)$$
This is the required conversion $(1)$ and we get in this way an encoding of Theorem \ref{refined}.

\subsection{Some variations}

 In \cite{Hurkens95}, Hurkens uses instead
 $$B = \Pi_{X:\Pos}(T~X\rightarrow X)\rightarrow T~X\eqno{(3)}$$
 He then develops a short paradox using this type $B$, but with a different intuition, which comes from
 Burali-Forti paradox. The  variation we present in this note starts instead
 from the remark that $T~A$ cannot be a retract
 of $A$. In \cite{coq94-2}, we also use this idea, but with a more complex use of partial equivalence relations,
in order to build a strong initial $T$-algebra from a
 weak initial $T$-algebra. This was following Reynolds' informal argument in \cite{Reynolds84}, 

The same argument from Theorem \ref{refined} can use the encoding
$(3)$ instead. We define then
$$\iota:\Pi_{X:\Pos}(T~X\rightarrow X)\rightarrow B \rightarrow X$$
by
 $$\iota~X~f~b = f~(b~X~f)$$
 and $\intro:T~B\rightarrow B$ by
 $$\intro~v~X~f = T~(\iota~f)~v$$
 We then have the choice for defining $\match:B\rightarrow T~B$. We can use
 $$\match = \iota~(T~B)~\intro$$
 as before. Maybe surprisingly, we also can use
 $$\match~b = b~B~\intro$$
 In both cases, we get the judgemental equality $\match\circ\intro = T~(\intro\circ\match)$ required for the use
 of Theorem \ref{refined}.

 \section{Computational behavior}

 For the paradox correspoding to Theorem \ref{simpl}, we have the following looping behavior
 with a term reducing to itself (in two steps) by \EMP{head linear reduction}

\medskip

\begin{tabular}{lcl}
  $l_2~p_0~l_2~l_1$ & $\rightarrow$ & $l_1~x_0~l_2~l_1$ \\
  & $\rightarrow$ & $l_2~p_0~l_2~l_1$ \\
  & $\rightarrow$ & $\dots$ \\
\end{tabular}

\medskip

\subsection{Family of looping combinators}

The paradox corresponding to Theorem \ref{refined} does not produce a term that reduces to itself
however

\medskip

\begin{tabular}{lcl}
  $l_0~p_0~l_2~l_1$ & $\rightarrow$ & $l_1~x_0~l_2~(s_2~p_0~l_1)$ \\
  & $\rightarrow$ & $l_2~p_0~(s_1~x_0~l_2)~(s_2~p_0~l_1)$ \\
  & $\rightarrow$ & $l_0~(p_0\circ\delta)~(s_1~x_0~l_2)~(s_2~p_0~l_1)$ \\
  & $\rightarrow$ & $s_2~p_0~l_1~x_0~(s_1~x_0~l_2)~(s_2~(p_0\circ\delta)~(s_2~p_0~l_1))$ \\
  & $\rightarrow$ & $l_1~(\delta~x_0)~(s_1~x_0~l_2)~(s_2~(p_0\circ\delta)~(s_2~p_0~l_1))$ \\      
  & $\rightarrow$ & $\dots$ \\
\end{tabular}

\medskip

Like for Hurkens' paradox however, we obtain a term that reduces to itself if we forget types in
abstraction  \cite{coq06}.

\medskip

In \cite{coq86}, I analysed another paradox, closer to Girard's original formulation 
(as was found out later by H. Herbelin and A. Miquel, a slight variation of this paradox
can be expressed in System $\U^{-}$.)
 At about the same time, A. Meyer and M. Reinholdt \cite{MeyerR86}, suggested a clever
 use of Girard's paradox for expressing a fixed-point combinator. While implementating this paradox
 \cite{coq86}, it was possible to check that, contrary to what \cite{MeyerR86} was hinting, the term representing this
 paradox was not reducing to itself\footnote{It would be interesting to go back to this paradox and check if it reduces to
 itself when  removing types in abstractions.}.
 A. Meyer found out then that it was however possible to use this paradox and produce a family
 of looping combinators instead, i.e. a term which has the same B\"ohm tree as
 one of a fixed-point combinator. A corollary, following \cite{MeyerR86}, is that type-checking is undecidable
 for $\type:\type$.

\subsection{Definitions and Head linear Reduction}

As discussed in \cite{Hurkens95}, using the notion of \EMP{definition} is essential, even for ``small'' terms,
for representing these paradoxes in an undertandable way. As was discovered in Automath \cite{deBruijn87},
in a type system with \EMP{dependent} types, one cannot reduce definitions to abstractions and applications like in simply typed
lambda calculus. Indeed, the representation of
$$\mathsf{let}~x:A = e_0~\mathsf{in}~e_1$$
by $(\lambda_{x:A}e_1)~e_0$ can be incorrect, since the definition $x:A=e_0$ can be used in the type-checking of $e_1$.

Furthermore, in order to understand the computational behavior of the paradox, the use of
\EMP{head linear reduction}, which plays an important role in \cite{deBruijn87}, is convenient.
This is what was done when presenting above the computational behavior of various paradoxes, with a
periodic behavior for the first example and a non periodic behavior for the paradox in $\U^{-}$.
This use may also be relevant for understanding large proofs.

\section*{Conclusion}

In this note, we presented a variation of Hurkens' paradox \cite{Hurkens95} and a paradox inspired by
Reynolds \cite{coq94-2}. This paradox can be seen as a refinement of the simple paradox presented in Theorem \ref{simpl}.
The problem is that in the encoding in $\U^{-}$, we don't get that $T~A$ is
a \EMP{judgmental} retract of $A$\footnote{This problem
was presented in \cite{coqP88} as one main motivation for the primitive introduction
of inductive definitions.}. It is possible however to still use a weaker judgemental equality and derive a relatively
simple paradox\footnote{We were not able however to refine in a similar way the paradox of trees
\cite{coq92-3}, to obtain a new paradox in $\U^{-}$.}.

\bibliography{synergy}
\bibliographystyle{plain}

\end{document}